# Breaking the Crowther Limit: Combining Depth-Sectioning and Tilt Tomography for High-Resolution, Wide-Field 3D Reconstructions

Robert Hovden[1], Peter Ercius[2], Yi Jiang[3], Deli Wang[4], Yingchao Yu[4], Héctor D. Abruña[4], Veit Elser[3], David A. Muller[1]

[1] School of Applied & Engineering Physics and Kavli Institute at Cornell for Nanoscale Science, Cornell University, Ithaca, NY 14853

[2] National Center for Electron Microscopy, Lawrence Berkeley National Laboratory, Berkeley, CA 94720

[3] Department of Physics, Cornell University, Ithaca, NY 14853

[4] Department of Chemistry and Chemical Biology, Cornell University, Ithaca, NY 14853

To date, high-resolution (< 1 nm) imaging of extended objects in three-dimensions (3D) has not been possible. A restriction known as the Crowther criterion forces a tradeoff between object size and resolution for 3D reconstructions by tomography. Further, the sub-Angstrom resolution of aberration-corrected electron microscopes is accompanied by a greatly diminished depth of field, causing regions of larger specimens (> 6 nm) to appear blurred or missing. Here we demonstrate a three-dimensional imaging method that overcomes both these limits by combining through-focal depth sectioning and traditional tilt-series tomography to reconstruct extended objects, with high-resolution, in all three dimensions. The large convergence angle in aberration corrected instruments now becomes a benefit and not a hindrance to higher quality reconstructions. A through-focal reconstruction over a 390 nm 3D carbon support containing over one hundred dealloyed and nanoporous PtCu catalyst particles revealed with sub-nanometer detail the extensive and connected interior pore structure that is created by the dealloying instability.

## Introduction

With electron beams smaller than the bond length of hydrogen, aberration-corrected scanning transmission electron microscopes (STEM) can image materials with resolutions below the shortest bond lengths in nature[1]. However, these atomic resolution images are only 2D projections of a specimen. In order to determine the full 3D structure, one must acquire a series of STEM images over a range of specimen tilts[2,3]. Unfortunately, the resolution of a 3D STEM tomogram is degraded by missing information that is a consequence of the restricted specimen tilt range and finite tilt increments. While reconstructions of small objects have been reported at atomic resolution[4,5], resolution places fundamental limitations to object-size. For larger objects (> 20 nm), the Crowther condition[1,6] discussed below typically limits volumetric resolutions of electron tomography to roughly 1 nm—twenty times worse than the best resolution in 2D projections. In addition, high-resolution (< 1 nm) tomography of general extended objects, to date, has not been possible with aberration-corrected instruments because the depth-of-field (5 – 10 nm) of aberration-corrected STEMs is smaller than most objects of interest today[7]. High-resolution 3D reconstruction of extended objects, i.e. those larger than the depth-of-field, requires collecting information beyond a traditional tilt series. Here we present through-focal tomography that combines depth sectioning and traditional tilt-tomography to reconstruct extended objects, with high-resolution, in three-dimensions. This combined approach fills in missing 3D information by acquiring a through-focal image series at each specimen tilt (Fig. 1)—decoupling the limiting Crowther relationship between 3D resolution and object size.

For traditional (S)TEM tomography the images at every tilt-angle must correspond to a perfect projection of the original object to fulfill the "projection requirement," and thus the entire specimen must always be in focus[2,3,7-10]. Reconstruction algorithms fail when images are not accurate projections of the specimen. According to the projection slice theorem, each projection corresponds to a plane of information in reciprocal space extending out to the lateral resolution limit of the microscope[4,11-13]. More tilts and a larger tilt range adds information and results in a higher quality reconstruction. Specimens with a needle shape geometry may permit the full 180° tilt rage[14]. However, practical limitations, such as the sample or stage geometry, often prevent full specimen rotation in the microscope—±70° is a common upper limit. This results in the infamous 'missing wedge' of information[14,15], as shown in Fig. 2b. Furthermore, it becomes unreasonable to tilt, align and focus over a large number of images, making it difficult to acquire more than ~150 tilt images.

The finite number of tilts causes smaller wedges of missing information between each tilt. These smaller missing wedges limit the resolution and 3D volume of a reconstructed object—as described by the Crowther condition: $d=\pi F/N$, where d is the reconstruction resolution for a field of view F, given N tilt projections[6]. Larger objects occupy a smaller volume in reciprocal space and require more tilts to adequately intersect and measure the information. An extreme example of undersampling is illustrated in Supplemental Figure 1, where two projections of a large crystalline sample are collected along planes that do not intersect any reciprocal lattice spots. It is not possible to reconstruct the object without collecting more experimental data such as additional projections that intersect the missing spots. Thus, reconstructing larger objects at higher resolution requires an increasing number of tilt images. With dose and practical time limitations, this ultimately results in a 3D resolution worse than the microscope's 2D resolution.

While improving the lateral resolution of the microscope—as accomplished with aberration-correctors—can sometimes improve the tomographic resolution, there are limitations and difficulties to consider. In high-resolution, aberration corrected STEM, the large convergence angle rapidly reduces the depth-of-field. Objects outside of the focal plane quickly become blurred and less defined. This has a detrimental effect on tomographic reconstructions, which assume a perfect projection image at every tilt. At best, this results in a degradation of resolution; at worst, objects will appear highly distorted or missing in the final reconstruction. A regime of extended objects beyond the depth-of-field are inaccessible to traditional electron tomography (Fig. 3). For reference, two representative aberration corrected machines in use today, Cornell's 100 keV NION UltraSTEM and NCEM's 300 keV FEI Titan TEAM I, have depths-of-field around 12 nm and 6.4 nm respectively (Supp. Fig. 2). For many real world objects, such as semiconductor devices or catalytic nanoparticles on a 3D support, it is not possible to image the entire specimen in focus—thus preventing high-resolution 3D imaging.

The limited depth of field can be overcome by depth sectioning in which a through-focal series of the specimen is acquired[16-19]. When imaging over a range of focal planes, object features lying on different focal planes appear in-focus at different times—indicating the presence of 3D information. However, a large amount of out-of-focus intensity remains in the images. One may be inclined to reconstruct the full 3D structure from a single through focal-image stack, however, any attempt will suffer from dramatic elongation artifacts[17,20]—for instance, a 5 nm particle can be elongated to 200 nm. This is a consequence of the missing cone of information in the contrast transfer function (CTF) of a through-focal series:

$$H(k_r, k_z) = \frac{2}{\pi \alpha_{max} k_r} \sqrt{1 - \left(\frac{k_r \lambda}{2\alpha_{max}} + \frac{k_z}{\alpha_{max} k_r}\right)^2}$$

where $\alpha_{max}$ is the electron beam convergence angle, $\lambda$ is the beam's wavelength, and $k_z$, $k_r$ are the reciprocal space cylindrical coordinates[21]. Fig. 2a shows the shape of a rotationally symmetric slice of a through-focal CTF, making apparent the large missing cone of information when depth sectioning an extended object. The low-frequency takeoff angle is described by the beam convergence angle, $\alpha_{max}$. In current high-resolution instruments the largest convergence angles are 30 mrad, corresponding to a takeoff angle in degrees of 1.7° with a wedge of information spanning 3.4° and a substantial 176.6° cone of missing information.

While a through-focal series lacks enough information for 3D reconstruction, it provides a sufficient amount of information for a 2D reconstruction—where a single 2D image shows all specimen features in focus. This extended depth of field (EDOF) technique[22], can be used to overcome the limitations of a small focal length in STEM[16]. EDOF insures that the information transfer limit is preserved over the entire specimen and allows for a traditional tomographic reconstruction. However, an EDOF approach discards information and does not take advantage of the continuum of 3D information provided in the through-focal series. Thus, we propose a through-focal tomography technique that uses the full 3D information contained within the focal-series stack.

Combining through-focal imaging with a traditional tilt series overcomes the limitations of aberration-corrected tomography, improves the reconstruction quality, and requires fewer tilts. The simple back projection used in traditional tomography does not include information provided by the probe shape (Fig. 2b). When the probe's depth-of-field is much larger than the object, as often is the case in uncorrected STEMs, the back projection is a reasonable approximation. However, in through-focal tomography, a through-focal image stack replaces each single projection image. This decouples the restricting relationship between object size and microscope resolution (convergence angle) allowing access to new 3D imaging regimes—illustrated in Fig 3.

The CTF of through-focal tomography is a superposition of through-focal CTFs rotated about the tilt axis. This utilizes the additional information of the propeller-shaped CTF at every tilt, not only overcoming the limitations of a small depth-of-field, but also filling in the smaller missing wedges of information present between each tilt (Fig 2c) and improving the quality of the 3D reconstruction (Fig. 6). When the increment tilt angle is smaller than the CTF takeoff angle, $\alpha_{max}$, the overlapping continuum of information collected lifts the Crowther tilt criterion and allows extended objects to be reconstructed at high resolution without increasing the number of tilts.

**Experiment**

To demonstrate through-focal tomography, we imaged porous dealloyed PtCu nanoparticles with TEAM I at the National Center for Electron Microscopy; a tool that provides the key attributes to best demonstrate the advantages of this technique. Its large convergence angle provides high lateral resolution (< 0.78 Å) and a small depth-of-field (~6 nm) at 300 kV accelerating voltage. Additionally, it's unique specimen stage provides a full tilt range (alpha = ±180°, beta = ±180°) for tomography; however shadowing from the TEM grid limited tilts from -68° to +71° along our chosen axis of rotation. The PtCu nanoparticles decorate a 3D Vulcan C support with an extended structure that far exceeds the microscope's depth of field—making it impossible to image multiple particles in-focus within a single field of view (Supp. Video 1).

The porous PtCu nanoparticles, with their potentially large electrochemically active surface areas were of interest for catalyzing the oxygen reduction reaction, a key step for improving the power density of polymer-electrolyte membrane fuel cells[23-27]. Since the pore structures of these particles are relatively unknown via 2-D projection images, tomography is necessary for a proper characterization. Traditional reconstructions of the pore structure may include, at most, a few particles in the field of view—providing only anecdotal observations. Using through-focal tomography, we simultaneously investigate the pore structure of many particles and their distribution on the Vulcan support.

Carbon supported Pt-Cu porous nanoparticles were synthesized via a three-step approach. First, the ordered intermetallic nanoparticles with 10wt% of Pt were prepared using an impregnation method. In a typical synthesis, 53.3 mg of $H_2PtCl_6·6H_2O$ and 41.4 mg of $CuCl_2$ were dissolved in ultrapure water, with 160.4 mg of Vulcan XC-72 carbon support dispersed into the liquid. After ultrasonic blending for 30 min, the suspension was heated under magnetic stirring to allow the solvent to evaporate and to form a smooth, thick slurry, which was dried in an oven at 60 °C. After being ground in an agate mortar, the resulting dark and free-flowing powder was heated in a tube furnace at 300 °C under flowing $H_2/N_2$ for 5 h. Second, the powder was cooled to room temperature under $N_2$. The as-prepared $Cu_3Pt/C$ powder was then annealed at 1000 °C under an $H_2$ atmosphere for 10 h to form an ordered intermetallic phase. Third, the carbon supported $Cu_3Pt$ ordered intermetallic nanoparticles were dealloyed by immersing the catalysts in 1 M HNO3 under magnetic stirring at 40°C for two days. The sample was then centrifuged and washed using deionized water until the pH value was close to 7. The clean PtCu nanoparticles with carbon support were suspended in ethanol and deposited onto a holey carbon ultrathin support film.

The tomography data was acquired over a 138° tilt range using a high angle annular dark field (HAADF) detector. The 30 mrad convergence angle provided a continuum of information in the through-focal CTF that spanned a ±1.72° wedge at low and medium frequencies. A 3° tilt increment was chosen to match the convergence angle. At every tilt a through-focal series was taken over ±250 nm defocus with 20 nm focal steps in order to insure all objects were imaged in focus. The microscope defocus steps are calibrated from a through-focal stack (Fig. 4). In total, the data amounted to 1645 images with 0.38 nm / pixel lateral resolution.

Before reconstruction, a five-dimensional alignment of the data was required: transverse x-y alignment, focal z-alignment, tilt axis rotation and shift. Each through-focal stack was first aligned in x-y to a fiduciary particle. Next a fiduciary particle was selected for alignment by identification of the best focus image in order to provide the focal z-alignment. The images within each focal stack were aligned using cross-correlation to correct for small amounts of drift during the acquisition. The data is reweighted in Fourier space by dividing with the microscope's CTF approximated by a 300 keV 30 mrad aberration free probe plus a Wiener constant of 5 times the max CTF value. After this light deconvolution, each through-focal stack is mapped onto a universal Fourier space by bilinear extrapolation, which distributes the complex value of an input point to its four nearest neighbors on the output Cartesian grid. The Fourier information of the object is a weighted average of points from all the through-focal stacks. The final reconstruction is obtained directly from the 3D inverse Fourier transform. The same Wiener filter was used in the traditional tilt series reconstruction for direct comparison.

**Results and Discussion**

The final through-focal tomographic reconstruction provided a high-resolution 3D image containing hundreds of PtCu nanoparticles on the support (Fig. 5, Supp. Video 2). The tomogram revealed the presence of interconnected pore structures throughout almost all the particles. However in some instances, smaller particles (< 5nm) were observed without pores—Figure 5 shows small particles with and without pores. The location, density, and size of pores appear independent of particle size**.** In all discernable cases, surface connected pore(s) were visible suggesting a surface instability during the dealloying process[29]. The dealloyed structure of these PtCu nanoparticles are reported to affect the catalytic activity in the oxygen reduction reaction of polymer-electrolyte membrane fuel cells[23].

The pore structure provided a good metric for the resolution of this novel technique. When compared to a traditional back projection, the improved quality of the through-focal tomography is clear—as seen in Figure 6. The depth resolution of an aberration corrected microscope, when combined with through-focal tomography, can provide improved resolution and artifact reduction even at atomic resolution (Supp. Fig. 5). However, the advantages of this technique become more obvious for extended objects larger than the microscope's depth of field. Unlike the traditional back projection, the through-focal tomogram correctly characterizes the pore structure in the PtCu nanoparticles over the entire extended carbon support (Fig. 5,6, Supp. Fig. 3,4). Additionally, the through-focal tilt-series preserves the high intensity of objects and reduces the fanning caustics that plague a simple back-projection (Fig. 6, Supp. Fig. 3). In the traditional back projection, particles far from the center focal plane appear blurred, distorted and their pore structure cannot be determined (Fig. 6d,f)—possibly leading to incorrect conclusions about particle morphology and certainly preventing any quantification. Furthermore, our traditional back projection reconstruction represented a best-case scenario. It was optimistically obtained from focal planes in the center of the object—something that is very difficult to execute during experimental acquisition without also acquiring a focal-series—information that is contained and used in the new approach, but discarded in the traditional approach.

The inevitable non-eucentricity of a specimen makes tilting, centering, and focusing a tedious process. By reducing the number of tilts, through-focal tomography has a practical advantage. Additionally, unlike a regular tilt-series, which requires time and often additional beam dose to determine the proper focus, through-focal tomography is relatively insensitive to the correct initial focus when the focal range is large enough to include all objects of interest. Often, automated focusing algorithms take and needlessly discard a focal series—information which through-focal tomography can use to improve any reconstruction. In this experiment 25 images were acquired per tilt, but 3 times fewer tilts were needed. Automated software with 8 images needed to autofocus would record the same number of images. Our experience with existing tilt-only acquisition software is 5-10 images are already used for focusing. In through-focal tomography this same information is incorporated in the reconstruction instead of being discarded, as in the traditional approach. For these reasons, this technique may also become useful in lower-resolution STEM tomography of extended objects such as biological specimens (withstanding dose limits) or semi-conductor devices where the specimen dimensions greatly exceed the microscope's depth-of-field.

The direct Fourier approach used here to reconstruct the through-focal data is simple, intuitive, and the artifacts are well understood. Non-linear and iterative reconstruction algorithms can make better use of the available information but will still fail for the incorrect, out-of-focus information of extended objects. For instance, low-frequency artifacts are reduced in traditional SIRT reconstructions, but pore structure is still unresolved, incorrectly defined, and a high level of noise is present in Fourier space (Figure 6e,f). Future work that combines through-focal tomography with advanced reconstruction methods would provide superior results. Recently, iterative optimization methods—such as total variance minimization, discrete tomography, and non-negativity with support constraints—utilize minimal prior knowledge of the object and have shown better estimation of the missing information[4,30-32]. These new approaches offer beneficial interpolation of missing information but are still subject to a finite depth of field and sampling limitations. These methods will fail when the projection requirement is not met and the information collected is incorrect—as can happen for large objects imaged in aberration corrected instruments.

Through-focal tomography is not a reconstruction algorithm or interpolation method. Through-focal tomography is a method for sampling the actual 3D information of an object, large or small, by collecting a swath of accurate 3D information at every tilt. New iterative reconstruction techniques should readily benefit from the additional and correct information collected in through-focal tomography.

## Conclusions

Here we have successfully demonstrated the first high-resolution 3D imaging of an extended object by developing a novel through-focal reconstruction technique. Utilizing the 30 mrad convergence angle of the 300 keV aberration-corrected TEAM I instrument, the high lateral and depth resolution provided ideal imaging conditions of porous PtCu nanoparticles lying on an extended (390 nm) 3D carbon support. While no single image was able to provide a high-resolution image of all particles in focus, a through-focal series acquired at every tilt overcame this limitation and provided high-frequency detail in the entire tomogram. Interconnected and surface-connected pore structures resulting from chemical treatment were observed in most PtCu nanoparticles. Traditional tomographic methods were not able to reconstruct the particles with sufficient detail to observe pores in all particles, demonstrating the utility of through-focal approaches for aberration corrected STEM tomography. Through-focal tomography allows high-resolution 3D reconstruction of extended objects while requiring fewer tilt angles and is less sensitive to initial focus determination. Experimental and simulated comparisons of through-focal tomography with traditional tomography methods reveal improved contrast and resolution throughout large fields of view.

## Acknowledgements

We acknowledge helpful discussions with Huolin Xin, Lena Fitting-Kourkoutis, Julia Mundy, and John Grazul. This work was supported by the Semiconductor Research Corporation, the Center for Nanoscale Systems at Cornell, an NSF NSEC (NSF #EEC-0117770, 0646547) and the Cornell Center for Materials Research, an NSF MRSEC (NSF DMR-1120296). Y. Jiang and V. Elser were supported by DOE grant DE-FG02-11ER16210. The experiments were performed at the National Center for Electron Microscopy, Lawrence Berkeley National Laboratory, which is supported by the U.S. Department of Energy under contract no. DE-AC02-05CH11231.

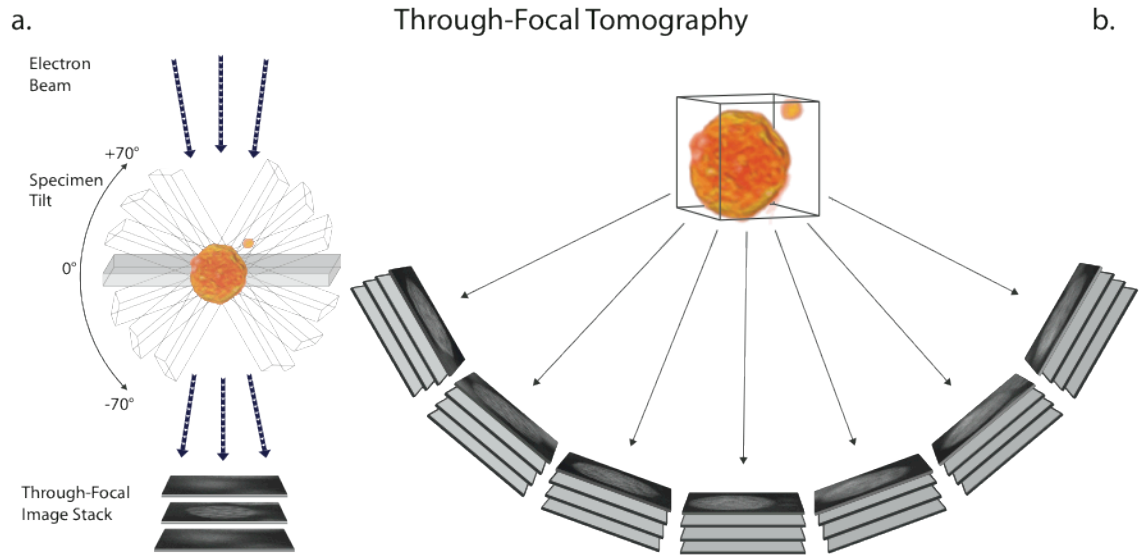

Figure 1 | Simplified diagram illustrating through-focal STEM tomography. a.) A stack of images at different focal planes is acquired at one specific tilt. This is repeated over a range of specimen tilts. b.) The through-focal image stacks acquired at every tilt angle contains lateral and depth information allowing for a high-resolution reconstruction of extended objects with fewer tilt angles.

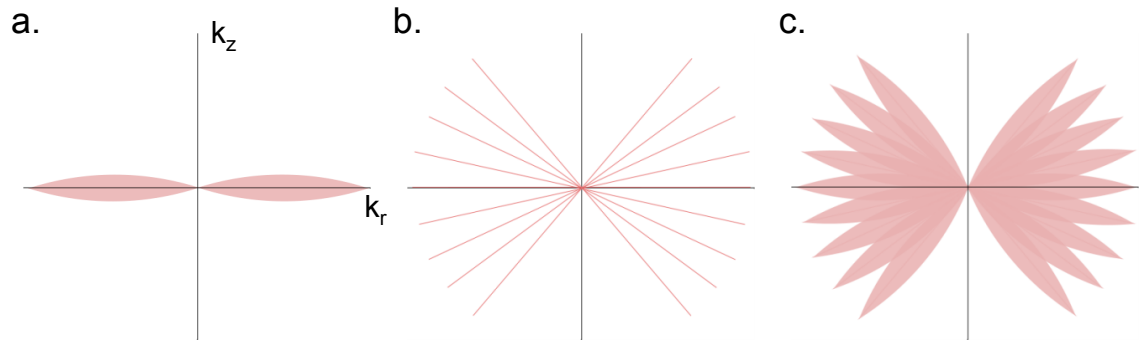

Figure 2 | Three diagrams show how various 3D reconstructions fill in information in reciprocal space: a) a slice of the rotationally-symmetric contrast transfer function (CTF) of an electron probe represents the information obtained from a through focal series. b) Slice through planar discs of information gained by traditional electron tomography. A finite number of tilts and a limited tilt range fail to sample much available information and there is still considerable missing information between the slices. c) Through-focal tomography combines approaches a) & b) to significantly increase the information sampled at each tilt.

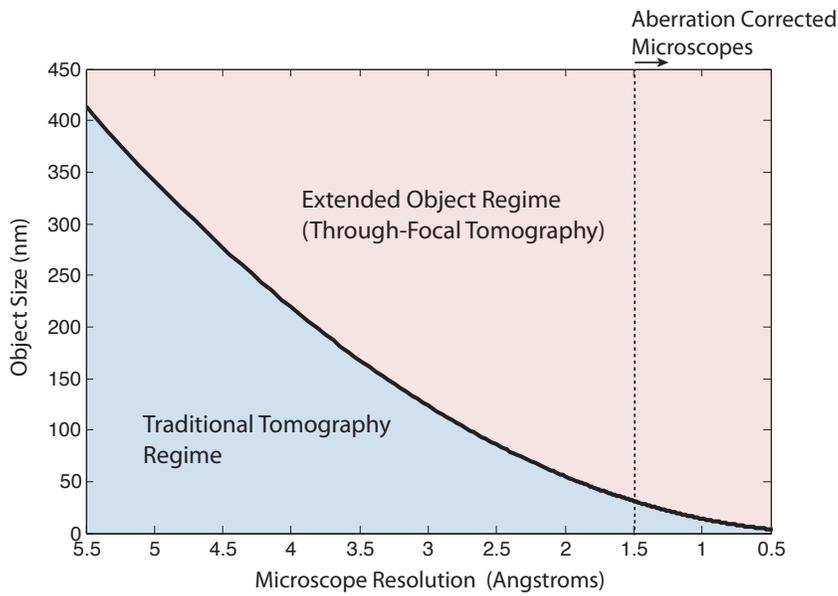

Figure 3 | Diagram shows the region accessible to traditional STEM tomography and the extended object regime which requires through-focal techniques in order to accurately reconstruct the 3D object. At higher resolutions, the extended objects can be small, c.a. 10 nm. The curve defining the extended object regime is based on the conservatively defined depth-of-field by Born & Wolf ($\lambda/\alpha^2$) and the Rayleigh Criterion for resolution ($0.61\ \lambda/\alpha$) at 300 keV.

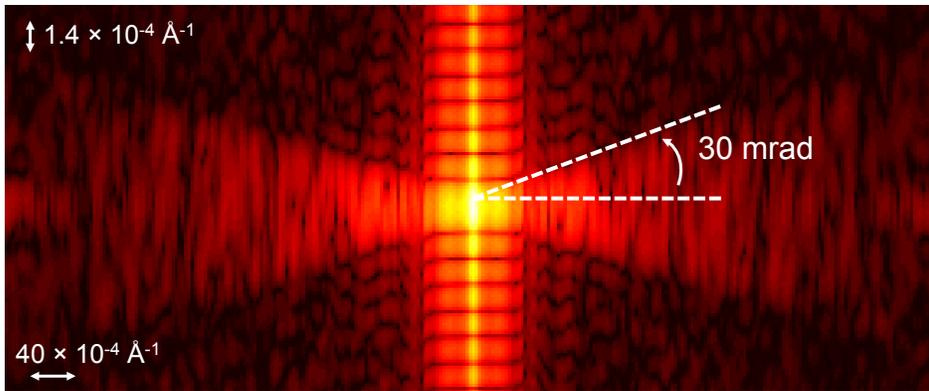

Figure 4 | Central slice through the 3D Fourier transform intensity of one through-focal image stack used in the tomographic reconstruction. The focus steps in the stack are calibrated in Fourier space using the known microscope convergence angle and field of view. Here, the convergence angle is 30 mrad with sampling in horizontal $dk_r = 2.8 \times 10^{-4}$ Å$^{-1}$ and calibrated beam direction $dk_z = 2.8 \times 10^{-4}$ Å$^{-1}$.

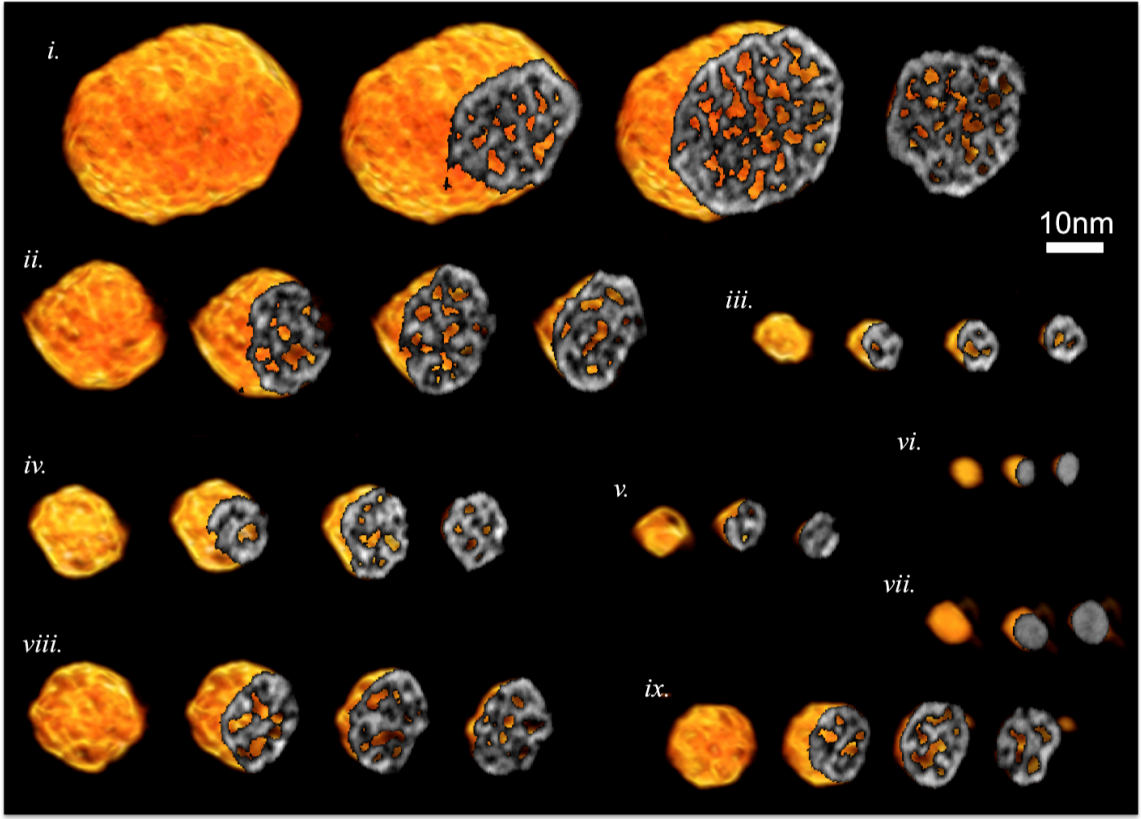

Figure 5 | High-resolution tomograms of many porous PtCu nanoparticles reconstructed using through-focal tomography. The particles were selected from a single reconstruction over a variety of positions across a 390nm support. Consecutive slices through each structure show the interconnected pore network of larger particles. However in some instances, smaller particles (< 5nm) were observed without pores.

## Through-Focal Tomography

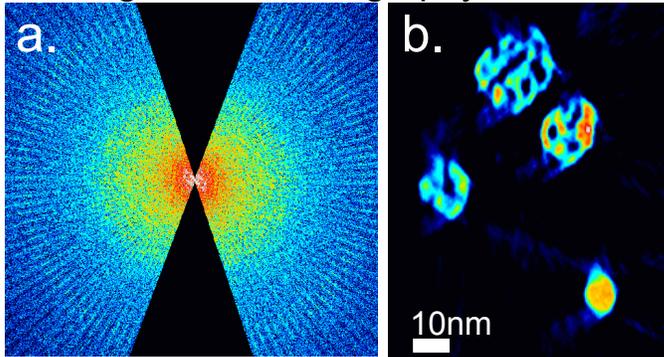

## Traditional Tomography – Backprojection

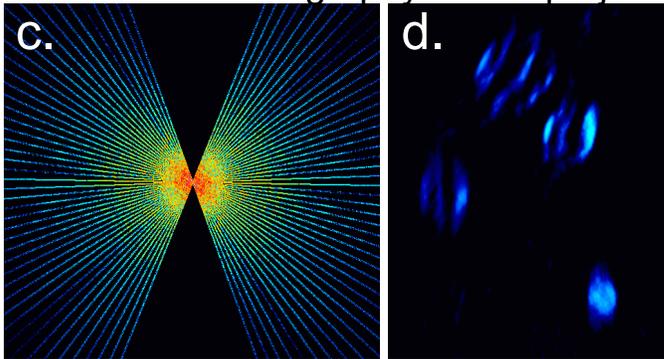

## Traditional Tomography – SIRT

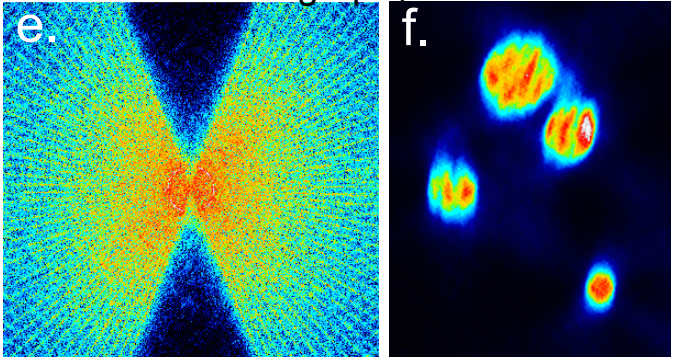

Figure 6 | Experimental comparison of through-focal tomography and traditional tomography of porous PtCu nanoparticles. In Fourier space (a) through-focal tomography acquires a continuum of information throughout the microscope's tilt range. In the real space reconstruction, dramatic differences in the techniques can be seen in the nanoparticle structure (a,d,f). A cross section of several nanoparticles shows the presence of pores in the through-focal reconstruction (b); features that are missing or too heavily distorted in the traditional reconstruction (d,f). Low frequency structure is improved in a nonlinear SIRT (f) reconstruction compared to the traditional backprojection of (d), but the pore structure is unresolved and still incorrectly defined. Artifacts in the 3D structure of the SIRT (f) result from the limited depth-of-field that accompanies aberration-corrected electron microscopes. This was overcome with through-focal tomography (c), which could be even further improved by nonlinear iterative methods. Intensity scale is low to high for blue to red coloring. Through-focal (a,b) and traditional-backprojection (c,d) images are placed on quantitatively comparable colorscales.

REFERENCES

1. Batson, P. E., Dellby, N. & Krivanek, O. L. Sub-ångstrom resolution using aberration corrected electron optics. *Nature* **418,** 617–620 (2002).
2. Midgley, P. A., Weyland, M., Thomas, J. M. & Johnson, B. F. G. Z-Contrast tomography: a technique in three-dimensional nanostructural analysis based on Rutherford scattering. *Chem. Commun.* 907–908 (2001). doi:10.1039/b101819c
3. Midgley, P. A. & Weyland, M. 3D electron microscopy in the physical sciences: the development of Z-contrast and EFTEM tomography. *Ultramicroscopy* **96,** 413–431 (2003).
4. Scott, M. C. *et al.* Electron tomography at 2.4-ångström resolution. *Nature* **483,** 444–447 (2012).
5. Zhu, C. *et al.* Towards three-dimensional structural determination of amorphous materials at atomic resolution. *Phys. Rev. B* **88,** 100201 (2013).
6. Crowther, R. A., DeRosier, D. J. & Klug, A. The Reconstruction of a Three-Dimensional Structure from Projections and its Application to Electron Microscopy. *Proceedings of the Royal Society A: Mathematical, Physical and Engineering Sciences* **317,** 319–340 (1970).
7. Hyun, J. K., Ercius, P. & Muller, D. A. Beam spreading and spatial resolution in thick organic specimens. *Ultramicroscopy* (2008).
8. DE ROSIER, D. J. & Klug, A. Reconstruction of Three Dimensional Structures from Electron Micrographs. *Nature* **217,** 130–134 (1968).
9. Hart, R. G. Electron Microscopy of Unstained Biological Material: The Polytropic Montage. *Science* **159,** 1464–1467 (1968).
10. Hoppe, W., Langer, R., Knesch, G. & Poppe, C. Protein-Kristallstrukturanalyse mit Elektronenstrahlen. *Naturwissenschaften* **55,** 333–336 (1968).
11. Bracewell, R. Strip Integration in Radio Astronomy. *Aust. J. Phys.* **9,** 198–217 (1956).
12. Wei, H. Fundamental limits of 'ankylography' due to dimensional deficiency. *Nature* **480,** E1–E1 (2011).
13. Wang, G., Yu, H., Cong, W. & Katsevich, A. Non-uniqueness and instability of 'ankylography'. *Nature* **480,** E2–E3 (2011).
14. Kawase, N., Kato, M., Nishioka, H. & Jinnai, H. Transmission electron microtomography without the 'missing wedge' for quantitative structural analysis. *Ultramicroscopy* **107,** 8–15 (2007).
15. Sugimori, H., Nishi, T. & Jinnai, H. Dual-Axis Electron Tomography for Three-Dimensional Observations of Polymeric Nanostructures. *Macromolecules* **38,** 10226–10233 (2005).
16. Hovden, R., Xin, H. L. & Muller, D. A. Extended depth of field for high-resolution scanning transmission electron microscopy. *Microscopy and microanalysis : the official journal of Microscopy Society of America, Microbeam Analysis Society, Microscopical Society of Canada* **17,** 75–80 (2011).
17. Nellist, P. D., Cosgriff, E. C., Behan, G. & Kirkland, A. I. Imaging Modes for Scanning Confocal Electron Microscopy in a Double Aberration-Corrected Transmission Electron Microscope. *Microscopy and microanalysis : the official journal of Microscopy Society of America, Microbeam Analysis Society, Microscopical Society of Canada* **14,** (2007).

18. Borisevich, A. Y., Lupini, A. R. & Pennycook, S. J. Depth sectioning with the aberration-corrected scanning transmission electron microscope. *Proc. Natl. Acad. Sci. U.S.A.* **103,** 3044–3048 (2006).
19. Behan, G., Cosgriff, E. C., Kirkland, A. I. & Nellist, P. D. Three-dimensional imaging by optical sectioning in the aberration-corrected scanning transmission electron microscope. *Philosophical Transactions of the Royal Society A: Mathematical, Physical and Engineering Sciences* **367,** 3825–3844 (2009).
20. Xin, H. L. & Muller, D. A. Aberration-corrected ADF-STEM depth sectioning and prospects for reliable 3D imaging in S/TEM. *Journal of electron microscopy* **58,** 157–165 (2009).
21. Intaraprasonk, V., Xin, H. L. & Muller, D. A. Analytic derivation of optimal imaging conditions for incoherent imaging in aberration-corrected electron microscopes. *Ultramicroscopy* **108,** 1454–1466 (2008).
22. Forster, B., Van De Ville, D., Berent, J., Sage, D. & Unser, M. Complex wavelets for extended depth-of-field: A new method for the fusion of multichannel microscopy images. *Microsc. Res. Tech.* **65,** 33–42 (2004).
23. Wang, D. *et al.* Tuning Oxygen Reduction Reaction Activity via Controllable Dealloying: A Model Study of Ordered Cu 3Pt/C Intermetallic Nanocatalysts. *Nano Lett.* **12,** 5230–5238 (2012).
24. Strasser, P. *et al.* Lattice-strain control of the activity in dealloyed core–shell fuel cell catalysts. *Nature Chem* **2,** 454–460 (2010).
25. Koh, S. & Strasser, P. Electrocatalysis on Bimetallic Surfaces: Modifying Catalytic Reactivity for Oxygen Reduction by Voltammetric Surface Dealloying. *J. Am. Chem. Soc.* **129,** 12624–12625 (2007).
26. Srivastava, R., Mani, P., Hahn, N. & Strasser, P. Efficient Oxygen Reduction Fuel Cell Electrocatalysis on Voltammetrically Dealloyed Pt–Cu–Co Nanoparticles. *Angew. Chem. Int. Ed.* **46,** 8988–8991 (2007).
27. Cui, C.-H., Li, H.-H., Liu, X.-J., Gao, M.-R. & Yu, S.-H. Surface Composition and Lattice Ordering-Controlled Activity and Durability of CuPt Electrocatalysts for Oxygen Reduction Reaction. *ACS Catal.* **2,** 916–924 (2012).
28. Wiener, N. Extrapolation, interpolation, and smoothing of stationary time series: with engineering applications. **9,** (1949).
29. Erlebacher, J., Aziz, M. J., Karma, A., Dimitrov, N. & Sieradzki, K. Evolution of nanoporosity in dealloying. *Nature* **410,** 450–453 (2001).
30. Saghi, Z. *et al.* Three-Dimensional Morphology of Iron Oxide Nanoparticles with Reactive Concave Surfaces. A Compressed Sensing-Electron Tomography (CS-ET) Approach. *Nano Lett.* **11,** 4666–4673 (2011).
31. Batenburg, K. J. *et al.* 3D imaging of nanomaterials by discrete tomography. *Ultramicroscopy* **109,** 730–740 (2009).
32. Bals, S. *et al.* Three-Dimensional Atomic Imaging of Colloidal Core–Shell Nanocrystals. *Nano Lett.* **11,** 3420–3424 (2011).